\documentclass[%
 reprint,%
 amsmath,%
 amssymb,%
 aps,%
 floatfix,
 superscriptaddress,
 prl]{revtex4-2}

\usepackage{graphicx}
\graphicspath{{./figs}}
\usepackage{mathtools}
\usepackage{amsmath, amssymb}
\usepackage{dsfont}
\usepackage{bm}
\usepackage[dvipsnames]{xcolor}
\usepackage[colorlinks,%
            bookmarksopen=false,%
            urlcolor=Blue,%
            citecolor=Blue,%
            linkcolor=Blue,%
            pdfencoding=auto,%
            psdextra]{hyperref}
\usepackage{physics}
\usepackage{upgreek}
\usepackage{bbm}
\usepackage[normalem]{ulem}
\usepackage{siunitx}
\usepackage{booktabs}
\usepackage{leftindex}

\newcommand{\sffrac}{\ensuremath{0.5}}

\begin{document}

\title{Assembling a Bose-Hubbard superfluid from tweezer-controlled single atoms}

\author{William J. Eckner}
\thanks{These authors contributed equally to this work.}
\affiliation{%
JILA, University of Colorado and National Institute of Standards and Technology,
and Department of Physics, University of Colorado, Boulder, Colorado 80309, USA
}%
\author{Theodor \surname{Lukin Yelin}}
\thanks{These authors contributed equally to this work.}
\affiliation{%
JILA, University of Colorado and National Institute of Standards and Technology,
and Department of Physics, University of Colorado, Boulder, Colorado 80309, USA
}%
\author{Alec Cao}
\thanks{These authors contributed equally to this work.}
\affiliation{%
JILA, University of Colorado and National Institute of Standards and Technology,
and Department of Physics, University of Colorado, Boulder, Colorado 80309, USA
}%
\author{Aaron W. Young}
\affiliation{%
JILA, University of Colorado and National Institute of Standards and Technology,
and Department of Physics, University of Colorado, Boulder, Colorado 80309, USA
}%
\author{Nelson \surname{Darkwah Oppong}}
\affiliation{%
JILA, University of Colorado and National Institute of Standards and Technology,
and Department of Physics, University of Colorado, Boulder, Colorado 80309, USA
}%
\affiliation{%
California Institute of Technology, Pasadena, California 91125, USA
}%

\author{Lode Pollet}
\affiliation{Department of Physics and Arnold Sommerfeld Center for Theoretical Physics (ASC), Ludwig-Maximilians-Universit{\"a}t M{\"u}nchen, Theresienstrasse 37, M{\"u}nchen D-80333, Germany}
\affiliation{Munich Center for Quantum Science and Technology (MCQST), Schellingstrasse 4, D-80799 M{\"u}nchen, Germany}

\author{Adam M. Kaufman}
\email[e-mail:$\,$]{adam.kaufman@colorado.edu}
\affiliation{%
JILA, University of Colorado and National Institute of Standards and Technology,
and Department of Physics, University of Colorado, Boulder, Colorado 80309, USA
}%

%\date{\today}

\begin{abstract}
    Quantum simulation relies on the preparation and control of low-entropy many-body systems to reveal the behavior of classically intractable models~\cite{feynman2018simulating}. 
    The development of new approaches for realizing such systems therefore represents a frontier in quantum science~\cite{browaeys2020many,gross2021quantum}.
    Here we experimentally demonstrate a new protocol for generating ultracold, itinerant many-body states in a tunnel-coupled two-dimensional optical lattice.
    We do this by adiabatically connecting a near-ground-state-cooled array of up to 50 single strontium-86 atoms with a Bose-Hubbard superfluid.
    Through comparison with finite-temperature quantum-Monte-Carlo calculations, we estimate that the entropy per particle of the prepared many-body states is approximately $2\,k_B$, and that the achieved temperatures are consistent with a significant superfluid fraction. This represents the first time that itinerant many-body systems have been prepared from rearranged atoms, opening the door to bottom-up assembly of a wide range of neutral-atom and molecular systems~\cite{gross2017quantum,gross2021quantum}.
\end{abstract}

\maketitle

Understanding the behavior of interacting many-body systems represents a frontier challenge in quantum information science~\cite{altman2021quantum}. While the large-scale entanglement inherent to these systems can challenge state-of-the-art theoretical methods~\cite{troyer2005computational,orus2019tensor},  synthetic quantum matter -- composed of neutral atoms, ions, or superconducting qubits~\cite{browaeys2020many,gross2021quantum,foss2024progress,kjaergaard2020superconducting} -- offers a testbed for exploring phenomena like quantum magnetism and transport, often with direct control of microscopic ingredients. Using neutral atoms, two prominent paths have emerged. In one, evaporatively-cooled quantum gases~\cite{anderson1995observation, demarco1999onset} are loaded into optical lattices, enabling quantum simulations of Bose- and Fermi-Hubbard models~\cite{greiner2002quantum,jordens2008mott,schneider2008metallic,sherson2010single, bakr2010probing,gross2017quantum}, with the potential to advance our understanding of high-temperature superconductivity and non-equilibrium dynamics~\cite{gross2017quantum,bourgund2025formation,xu2025neutral}. In another,  a many-body system is assembled atom-by-atom in highly-reconfigurable arrays of optical tweezers~\cite{endres2016atom,barredo2016atom,kumar2018sorting}, enabling ground-breaking studies of, as examples, quantum magnetism, frustration, and symmetry breaking~\cite{bernien2017probing, semeghini2021probing,browaeys2020many,scholl2021quantum, chen2023continuous}. 

These two paths beg the question of whether the bottom-up control arising in tweezer systems might be applied to the preparation of many-body ground states of itinerant systems. Indeed, this question has already motivated experiments that combine the programmability of tweezer arrays with the physics of itinerant bosons and fermions prepared by evaporative cooling~\cite{murmann2015two, yan2022two}, as well as demonstrations of two-particle Bose-Hubbard dynamics initialized by laser cooling~\cite{kaufman2014two, kaufman2015entangling}.
While Bose-Hubbard superfluids are typically prepared by loading an evaporatively-cooled gas into an optical lattice, it was proposed two decades ago that they might be assembled by placing atoms at the sites of an optical lattice and laser cooling each in parallel to the ground state of their respective lattice sites~\cite{olshanii2002producing}. By ramping down the lattice intensity adiabatically, it was theoretically shown that, with sufficient control,  this initial product state of Fock states should condense, thereby realizing quantum gas assembly~\cite{olshanii2002producing}. 
Analogous concepts have been explored in a chain of superconducting qubits~\cite{saxberg2022disorder}. 

Quantum gas assembly via atomic rearrangement and laser cooling has a number of potential advantages. It could be used to prepare in-lattice many-body ground states, or continuum states, for atoms and molecules whose collisional properties hinder evaporation. 
Further, in the quest to use quantum simulation to unravel the low-temperature physics of the Fermi-Hubbard model, new tools capable of approaching zero entropy through atomic rearrangement and laser cooling might establish versatile and potentially more effective pathways to state preparation~\cite{gross2017quantum,bourgund2025formation,xu2025neutral}. 
From a practical standpoint, reaching degeneracy assisted or purely by laser cooling has the advantage that its speed is determined by a transition linewidth rather than a collisional rate~\cite{stellmer2013laser, hu2017creation, xin2025fast}, which is a key consideration for measuring quantities like entanglement~\cite{moura2004multipartite, daley2012measuring, islam2015measuring, kaufman2016quantum, elben2023randomized} or correlation functions, both demanding high statistics~\cite{mazurenko2017cold, bourgund2025formation}. 
Lastly, as tweezer-based rearrangement produces Fock states, quantum gas assembly probes a novel regime of many-body physics and quantum optics, in which many-body phase coherence and well-defined atom number coexist, a tension that has been considered since the early days of Bose-Einstein condensation~\cite{castin1997relative,andrews1997interference}.

\begin{figure*}
    \includegraphics[width=\textwidth]{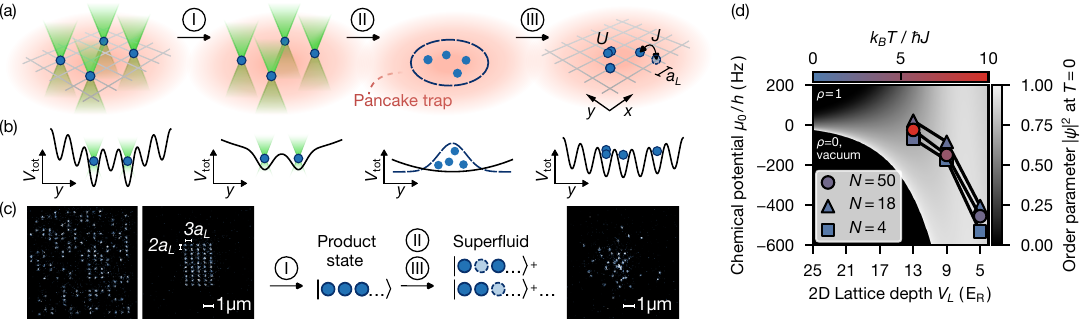}
    \caption{\label{fig:1}%
    \textbf{Tweezer prepared states of interacting atoms.}
    \textbf{(a)}~Schematic illustration of the experimental protocol. 
    Roman numerals indicate and label a corresponding series of ramps that take potentials from one panel to the next.
    ${{}^\mathrm{86}\mathrm{Sr}}$ atoms (blue circles) near their motional ground state are initially trapped in a combination of optical tweezers (green double cone, wavelength $515\,\rm{nm}$) and a two-dimensional (2D) square optical lattice (gray lines, with lattice constant $a_L$). 
    A highly anisotropic `pancake trap' (light-red shading) is realized by overlapping a single plane of a one-dimensional (1D) optical lattice along the axis $\hat{z} = \hat{x} \times \hat{y}$~\cite{young2023programmable}.
    $J$ and $U$ correspond to the hopping and interaction parameters, respectively, in the Bose-Hubbard model, which describes the system when the 2D lattice is in the tight-binding regime.
    \textbf{(b)}~Black curves (not drawn to scale) depict 1D cuts through the total potentials shown in (a), including the lattice, tweezers, and pancake trap. 
    \textbf{(c)}~From left to right: a single-shot image of the random initial fill for atoms in the array; a single-shot image of atoms after successful rearrangement; arrows representing adiabatic ramps of the optical potentials, and ket-vectors indicating the adiabatic transformation from a classical product state to a superfluid; finally, a single shot image of the itinerant state in the 2D lattice.
    \textbf{(d)}~Phase diagram for the Bose-Hubbard model. 
    Values from a zero-temperature mean-field calculation (see Methods) of the norm-squared order parameter at the center of the trap (where we take $i=0$) $|\psi|^2 = |\langle \hat{b}_0 \rangle|^2$ as a function of the chemical potential $\mu_0$ and lattice depth $V_L$ are indicated by the black-gray-white colorbar. 
    Blue, purple, and red points are the locations in the phase diagram corresponding to experimental data, and the color represents the temperature of the system at each point.
    The black region labeled $\rho = 0$ corresponds to vacuum states, and the region labeled $\rho = 1$ corresponds to unit-filling Mott-insulator states.}
\end{figure*}

In this work, we report on the assembly of low-temperature states of the two-dimensional (2D) Bose-Hubbard model.
This approach does not rely on evaporation and allows for a well-defined atom number.
The 2D Bose-Hubbard model exhibits a finite-temperature superfluid transition, described by the celebrated Berezinskii–Kosterlitz–Thouless
(BKT) mechanism~\cite{kosterlitz1973ordering, sunami2025detecting}. 
In our system, the situation is also modified by finite-size effects from the trapping potential, which enable the possibility of Bose condensation in two dimensions~\cite{hadzibabic2008trapped}.
In time-of-flight (TOF) expansions from the lattice, we find that the tweezer-prepared states exhibit diffraction peaks consistent with phase coherence (see Methods) and quantum degeneracy~\cite{greiner2002quantum}.
Pursuing more quantitative benchmarks, thermometry is performed by comparing measured densities to numerically exact finite-temperature quantum-Monte-Carlo (QMC) calculations.
These comparisons also allow us to infer the presence of superfluidity, confirm that the temperatures are consistent with observed TOF expansions, and estimate that data with $N=18$ atoms correspond to a trap-averaged (trap-center) entropy per particle $S/N$ of about $2.0\,k_B$ ($0.5\,k_B$), where $k_B$ is Boltzmann's constant.

\begin{figure*}
    \includegraphics[width=\textwidth]{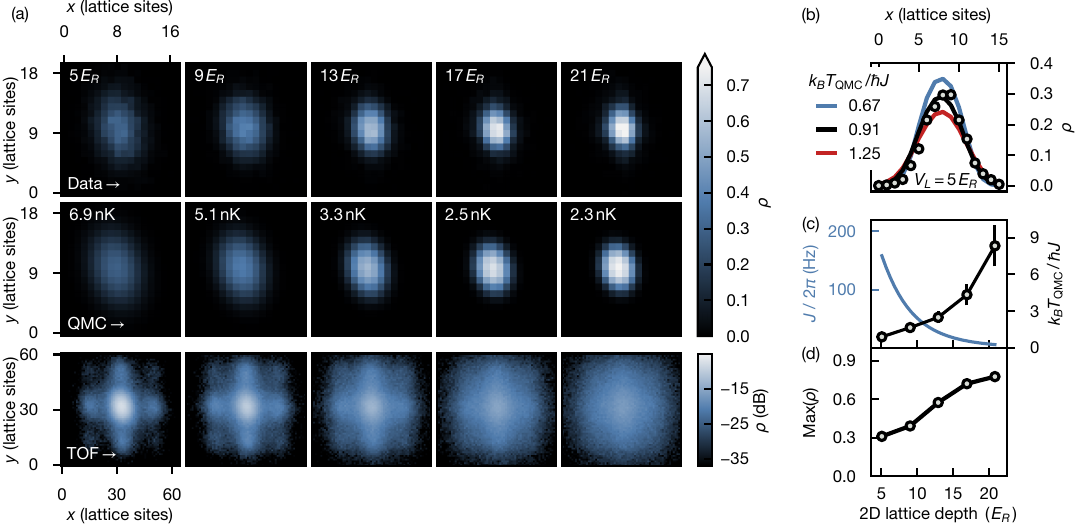}
    \caption{\label{fig:2}%
    \textbf{Many-body thermometry by comparison to quantum Monte Carlo.}
    \textbf{(a)}~\textit{Row} 1: Mean fluorescence images of lattice-prepared states with $N=18$ and lattice depths indicated in units of $E_R$. The density $\rho(x,y)$ is parity-projected by the imaging protocol.
    \textit{Row} 2: Quantum-Monte-Carlo (QMC) simulations of the parity-projected density $\rho(x,y)$ at temperatures quoted in the top left of each image, and $N=18$. Lattice depths match those stated in the corresponding panels in the first row.
    \textit{Row} 3: Mean fluorescence images of atoms after a time of flight (TOF) performed by quenching the 2D lattice off for $1.5\,\rm{ms}$, and then quenching it back on to pin atoms in deep potential wells for imaging.
    \textbf{(b)}~Gray circles with black edges show the parity projected density $\rho(x,y)$ for the central row of data with $N=18, V_L = 5.1(1)\,E_R$ [top left panel of part (a)].
    Solid lines come from QMC calculations at temperatures quoted in the legend.
    \textbf{(c)}~\textit{Right axis}: The thermal energy $k_B T$ for temperatures of the QMC results shown in row 2 of (a). \textit{Left axis:} The hopping parameter $J$ plotted continuously versus lattice depth.
    \textbf{(d)}~The maximum value of $\rho$ for the data in the top row of (a) is shown versus 2D lattice depth.}
\end{figure*}

Fig.~\ref{fig:1}(a) schematically illustrates the experiment. Ensembles of ${}^{86}\rm{Sr}$ atoms are trapped in a combination of a programmable 2D tweezer array, a square 2D optical lattice with spacing $a_{L} = 575\, \rm{nm}$, and a single plane of a one-dimensional (1D) optical lattice, referred to as the `pancake trap.' The pancake trap provides tight confinement along the vertical direction, and has Gaussian waists of about $43\,a_L$ and $63 \, a_L$ in the $x$-$y$ plane (see Methods for details)~\cite{young2022tweezer, eckner2023realizing, cao2024multi}. 
Atoms are initially captured from a narrow-line magneto-optical trap (MOT) into a $16\times24$ tweezer array, in which each tweezer has a roughly $50\,\%$ chance of containing one or zero atoms.
After transferring atoms into the 2D lattice, we take a low-loss image of the stochastically loaded array, and then tweezer-rearrange atoms into a rectangular sublattice of target sites with dimension $L_x \times L_y = N$ and spacings of $3\,a_{L}$ and $2\,a_{L}$ along the $x$ and $y$ axes respectively.
For the experiments in this work, $N \in \{4, 18, 50 \}$, each with dimension $(L_x,  L_y) \in \{(2,2), (3,6), (5,10) \}$, respectively.
After rearrangement, we take a second image to determine whether the target pattern was prepared successfully.
Single-shot images of stochastically filled lattice sites, as well as atoms rearranged into an $N=50$ sublattice are shown in Fig.~\ref{fig:1}(c).
We post-select $N=4$ and $N=18$ ($N=50$) data on perfect rearrangement ($\ge 90\, \%$ rearrangement), meaning that $N$ atoms ($\ge 0.9 \times N$ atoms) are detected within the target sites, and no atoms are detected anywhere else.

After rearrangement and laser cooling, the $515 \, \rm{nm}$ tweezers are ramped up, and then the 2D lattice is ramped off, as represented in Fig.~\ref{fig:1}(a) `I.'
${}^{86}\rm{Sr}$ has a large $s$-wave scattering length of $810.6\,a_0$~\cite{aman2018photoassociative}, where $a_0$ is the Bohr radius.
Therefore, when the tweezers are deep, these repulsive interactions create an energy penalty for having multiple atoms in the same site and result in a many-particle ground state that has significant overlap with a product state of a single atom per tweezer. 
Next, Fig.~\ref{fig:1}(a) `II' indicates the step in which this ensemble of Fock states is adiabatically transformed into a harmonically trapped Bose gas. In particular, the unit-filled tweezer array is ramped off, and atoms are transferred into the pancake trap, which has angular trap frequencies of $2\pi \times \,94.8(9)\,\rm{Hz}$, $2\pi \times \,63.2(9)\,\rm{Hz}$ in the $x,y$ plane, and $2\pi \times \,3.95(1)\,\rm{kHz}$ along the $z$ direction. 
Keeping the 2D lattice off during step `II' is intended to maximize kinetic energy scales, thermalization rates, and adiabaticity, however we have observed that having the lattice at a depth of $V_L \approx 5\,E_R$ during the tweezer ramps does not degrade the quality of state preparation in complementary measurements.
Here, $E_R = \pi^2 \hbar^2 / (2 m  a_{L}^2) = 2\pi \hbar \times 1.75 \, {\rm{kHz}}$ is the lattice recoil energy, $m$ is the atomic mass of ${}^{86}\rm{Sr}$, and $\hbar$ is the reduced Planck constant.
Finally, after the tweezers are fully turned off, the 2D lattice is ramped up to a variable trap depth of $V_L > 0$, as indicated by Fig.~\ref{fig:1}(a) step `III.'
We confirm that this protocol does not cause any measurable interband heating by performing sideband thermometry on the ${}^1{\rm{S}}_0\leftrightarrow{}^3{\rm{P}}_1$ transition (see Methods section Sideband cooling and Extended Data Fig.~\ref{fig:1ED})

\begin{figure*}
    \includegraphics[width=\textwidth]{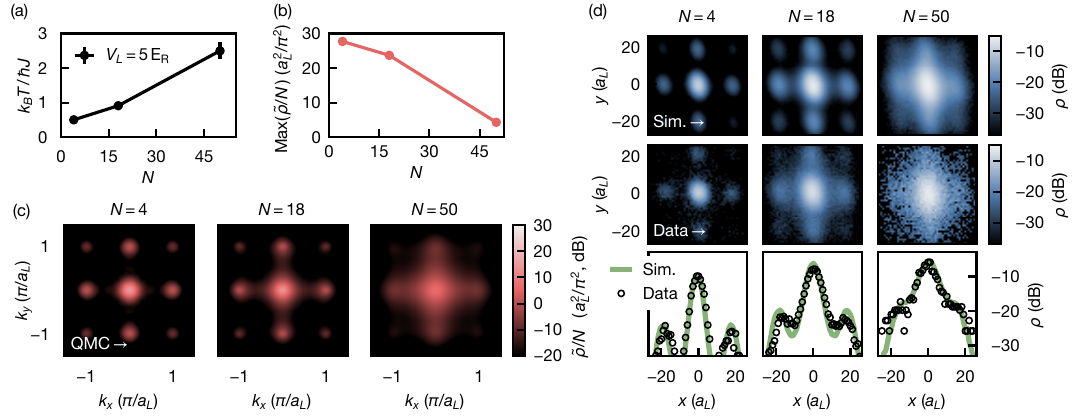}
    \caption{\label{fig:3}%
    \textbf{Temperature and time-of-flight expansions versus system size.}
    All data and simulations in this figure are performed with a lattice depth of $V_L = 5.1\,E_R$. 
    \textbf{(a)}~Temperature extracted by comparison to QMC calculations (as in Fig.~\ref{fig:2}) versus $N$. 
    \textbf{(b)}~peak values of the real-space momentum distributions shown in (c).
    \textbf{(c)}~QMC calculations of the momentum distributions for the $N$ and $T$ in (a).
    \textbf{(d)}~\textit{Row 1:} Using the momentum distributions in (c), we perform a semi-classical calculation (described in the main text) to simulate a $1.5\,\rm{ms}$ time-of-flight (TOF) expansion for the system in the pancake trap.
    The simulation results are shown in the top row.
    \textit{Row 2} displays mean fluorescence images for a TOF expansion of $1.5\,\rm{ms}$ in the pancake trap.
    \textit{Row 3} compares the parity-projected density $\rho(x,y)$ for the central rows of lattice sites.
    The open black circles correspond to experimental data, and the green curve is the result of the semi-classical simulation. 
    }
\end{figure*}

Once the 2D lattice reaches its final depth, the system can be described by the Bose-Hubbard model with nearest-neighbor tunneling $J$ in both directions of the square lattice, and on-site interaction strength $U$. 
The corresponding Hamiltonian is given by
\begin{align}
    \frac{\hat{H}}{\hbar} &= - J \sum_{\langle i,j\rangle} \left( \hat{b}_i^\dagger \hat{b}_j + \rm{h.c.} \right) \nonumber \\ & \ \ \ \ +  \sum_i \left[ \frac{U}{2} \hat{n}_i \left( \hat{n}_i - 1 \right) - \frac{\mu_0 + \Delta\mu_i}{\hbar} \hat{n}_i \right],
\end{align}
where $\langle i, j \rangle$ indicates a summation over pairs of indices $(i,j)$ that correspond to adjacent sites in the lattice, $\hat{b}_i^\dagger$ ($\hat{b}_i$) is the boson creation (annihilation) operator at site $i$, $\hat{n}_i =\hat{b}_i^\dagger \hat{b}_i $, $\mu_0 + \Delta\mu_i$ is the chemical potential at site $i$, and $\mu_0$ is the chemical potential at the center of the trap.
In this work, we operate at five different 2D-lattice depths with $V_L / E_R \in \left\{5.1(1), 9.0(1), 12.9(1), 16.9(2), 20.8(3) \right\}$.
Projective measurements are performed after quickly ramping up the lattice to pin atoms in place.
Following these ramps, low-loss fluorescence images enable single-site-resolved atomic detection~\cite{young2024atomic}.
This imaging is not fully number resolving and rather resolves parity: when there is an odd (even) number of atoms on a lattice site, imaging detects the presence of one (zero) atoms due to light-assisted collisions~\cite{DePue1999unity,sherson2010single, bakr2010probing}.
This parity-projected density for the lattice site at position $(x,y)$ will be denoted $\rho(x,y)$.
Because the shape of the distribution $\rho(x,y)$ is sensitive to the temperature $T$, we can perform thermometry by estimating $T$ to be the value that leads to a QMC-calculated distribution for $\rho(x,y)$ that best reflects the data.
Fig.~\ref{fig:2}(b) shows one example of this procedure, and inferred temperatures for $N=18$ data at all of the lattice depths are shown in Fig.~\ref{fig:2}(c).

An important question is whether the assembly protocol successfully prepares a superfluid state in our system.
This can be investigated with a QMC calculation for a homogeneous, $10 \times 10$-lattice-site system with atomic density $n_{\rm hom}$ equal to the central density for a given dataset.
As the superfluid fraction is expected to be larger at lower lattice depths, we focus on a homogeneous system with  $n_{\rm hom} = 0.324$, corresponding to the density of $N = 18$ data with $V_L = 5.1(1)\,E_R$.
At this 2D-lattice depth, we have $J = 2\pi\times 158(3) \, \rm{Hz}$, $U = 2\pi\times 6.2 \, J$, and, for $N=18$ data, $k_B T  = 0.9(1) \, \hbar J$.
In the $10\times10$-site homogeneous system with $n_{\rm hom} = 0.324$, we calculate a superfluid fraction of $f_s \approx \sffrac$~(see Methods).
We note that for fixed density and temperature, $f_s$ decreases with system size \cite{carrasquilla2013superfluid}.
Because the $10 \times 10$-sites is significantly larger than the experimental density distribution, we use this result to infer an appreciable ($ > \sffrac$) trap-center superfluid fraction (see Methods).
When the homogeneous QMC simulation is scaled to sizes beyond that of the trapped system, the calculated superfluid fraction goes down, eventually reaching zero by a system size of $64 \times 64$.
This indicates that the temperature $T = 0.9(1) \, \hbar J / k_B$ is above the threshold for superfluidity in the thermodynamic limit, and finite-size effects play an important role.
Indeed, we find using QMC that $T_c \approx 0.685 \, \hbar J / k_B$ for the interaction strengths $U$ explored in this work, which is slightly below our inferred experimental temperatures, and, notably, near to the critical temperature found for hard-core bosons~\cite{carrasquilla2013superfluid}.

Next, we illustrate the phase coherence (see Methods) in the system with 2D time-of-flight (TOF) expansions, performed by quenching off the 2D lattice after state preparation.
During the TOF, the pancake trap is left on, which keeps atoms in the focal plane of the imaging system.
After $1.5\,\rm{ms}$, the 2D lattice is quenched back on, once again pinning atoms in deep potential wells~\cite{bakr2010probing}.
The atoms are then sideband cooled and imaged.
Mean images after TOF expansions with $N=18$ atoms are shown in the bottom row of Fig.~\ref{fig:2}(a).
TOF expansions from the shallowest lattice depth of $V_L = 5.1(1)\,E_R$ exhibit diffraction peaks, consistent with the presence of spatial coherence and superfluidity~\cite{greiner2001exploring, greiner2002collapse, greiner2002quantum}. 
When the same procedure is repeated after ramping to deeper lattice depths, we see these peaks disappear, characteristic of a transition to a state with reduced spatial coherence.
 
In a $T=0$ Bose-Hubbard system with unity filling, increasing $J/U$ by ramping to larger $V_L$ leads to the well-known superfluid-to-Mott-insulator crossover~\cite{jaksch1998cold, greiner2002quantum,bakr2010probing, sherson2010single}. 
For our atom numbers and external confinement, the Mott-insulator (MI) transition occurs at values of $J/U$ that are below the critical value $(J/U)_{\rm crit.} \approx 0.06$ for larger, unit-filling states in 2D~\cite{mahmud2011finite}.
For example, QMC simulations with $T \approx 0$ and 18 atoms (50 atoms) demonstrate the absence of a MI state even when $J/U = 0.005$ ($J/U = 0.013$), corresponding to $V_L = 17 \, E_R$ ($V_L = 13 \, E_R$), while we expect a MI by $V_L \approx 21 \, E_R$.
Due to the higher lattice depths and corresponding low tunneling rates ($J \lesssim 6 \, \rm{Hz}$) near the transition, we hypothesize that we cannot ramp adiabatically into the MI phase. 
Hence, the reduction in phase coherence observed at higher lattice depths is the result of leaving the superfluid phase and entering a normal phase, rather than transitioning to a MI state.
To qualitatively illustrate this point, Fig.~\ref{fig:1}(d) shows how the QMC-inferred values of $\mu_0$ are located below the lower boundary of the unit-filling (labeled $\rho=1$) Mott lobe in the $\mu_0$-$V_L$ (mean-field) phase diagram.

In Fig.~\ref{fig:3}, we investigate the momentum distributions for systems with $N = 4, 18, 50$ and $V_L = 5.1(1)\,E_R$.
The data in Fig.~\ref{fig:3}(a) correspond to the temperatures inferred by comparing $\rho(x,y)$ to finite-temperature QMC results, as in Fig.~\ref{fig:2}.
The normalized momentum distributions that correspond to these temperatures can be calculated (see Methods), and are shown in Fig.~\ref{fig:3}(b) and Fig.~\ref{fig:3}(c), which represent the peak momentum densities and full momentum distributions, respectively, versus $N$.
In an ideal, free-space TOF expansion, the final position distribution would be a measure of the momentum distribution of the system, which we denote $\tilde{\rho}(k_x,k_y)$, where $k_x, k_y$ are the coordinates of the momentum wave vector.
However, in order to keep atoms in the imaging plane, TOF experiments are performed with the pancake trap left on, thereby subjecting atoms to the effects of harmonic confinement.
Nevertheless, we can still compare QMC-generated momentum distributions, shown in Fig.~\ref{fig:3}(c), to measured TOF patterns, shown in Fig.~\ref{fig:3}(d), by simulating some of the experimental details of the expansion with a semi-classical procedure (see Methods).
The bottom row in Fig.~\ref{fig:3}(d) shows that the semiclassical simulations match the experimental data well, providing further confidence that the QMC-calculated distributions are representative of the underlying momentum distribution.
Furthermore, all of the momentum distributions exhibit diffraction peaks.
Notably, the contrast of the peaks and value of the momentum density decrease with atom number.
Both a higher $T$ and the spread of atoms over lattice sites with higher values of $\Delta\mu_i$ could be responsible for this reduction.

\begin{figure}
    \includegraphics[width=\columnwidth]{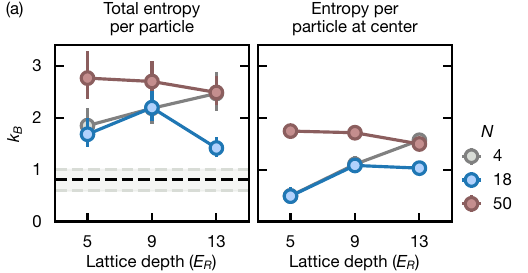}
    \caption{\label{fig:4}%
    \textbf{Entropy.}
    \textbf{(a)}~Curves correspond to the total entropy $S$ divided by $N$ as a function of lattice depth, and for temperatures in Fig.~\ref{fig:3}(a). 
    The black dashed line indicates the expected $S/N$ for initially laser-cooled atoms, and the vertical width of the gray shaded region corresponds to 2$\sigma$ uncertainty in this estimate (see Methods).
    \textbf{(b)}~The inferred entropy per particle at the center of the harmonic confinement -- denoted $(S/N)_{\rm cent.}$ in the main text -- plotted as a function of lattice depth.
    For both (a) and (b), gray, blue, and red points represent $N=4, 18,$ and $50$ respectively. 
    }
\end{figure}

Next, the total entropy $S$ is calculated by integrating the thermodynamic relation
\begin{align}
    dS = \beta \, d\langle E \rangle  - \beta \sum_i \left( \mu_0 + \Delta \mu_i \right) \,d\langle n_i \rangle,
\end{align}
where $\langle n_i \rangle$ and $\langle E \rangle$ are the QMC-calculated mean atom-number at site $i$ and total energy, respectively, for a given value of the inverse temperature $\beta$ at fixed chemical potential $\mu_0$.
The limits of integration are set by the inferred value of $\beta$ for the data (on the low-temperature end), and a high-temperature cutoff, where the entropy can be calculated with a high-temperature series expansion (see Methods).
Fig \ref{fig:4}(a) shows the inferred values for $S/N$ versus atom number and lattice depth.
To further understand the prepared states, Fig.~\ref{fig:4}(b) also displays the inferred entropy per particle at the center of the trap, denoted $(S/N)_{\rm cent.}$ (see Methods), and we find that $(S/N)_{\rm cent.} < S/N$.
Fig \ref{fig:4}(a) also indicates the expected entropy contribution due to imperfect sideband cooling, which is about half the total $S/N$ for $N=4, 18$. 

In summary, we have demonstrated assembly of ultracold many-body states from rearranged arrays of laser-cooled atoms in a hybrid tweezer-lattice architecture. 
The reported observations suggest several paths toward improved scalability and lower entropy in future work, and, in particular, reaching temperatures below the (superfluid) critical temperature in the thermodynamic limit, which is a prerequisite to system size scaling. 
For example, the disparity between the entropy per particle before and after many-body state preparation -- as illustrated in Fig \ref{fig:4}(a) -- might be explained by heating from intensity and laser-beam-pointing noise in the optical potentials. 
Minimizing the amount of atomic transport during adiabatic ramps, by improved matching of the equilibrium density distribution among each step of the preparation sequence, would also likely lead to improved performance.
In this direction, rearranging atoms into nearest-neighbor lattice sites, followed by sideband cooling would enable the direct assembly of a unit-filled Mott-insulator state in the lattice.
The limit on performance would then likely be set by sideband cooling, which can be improved with the use of larger trap frequencies as well as error-detection schemes such as erasure cooling~\cite{shaw2025erasure}.

This work also paves the way for future applications of tweezer-control techniques to the observation of dynamical superfluid properties~\cite{desbuquois2012superfluid} and their onset versus particle number, which could further elucidate the transition from the few to many-body regime~\cite{wenz2013few, bayha2020observing}.
Looking toward a broader set of itinerant systems, operation with the fermionic isotope ${}^{87}\rm{Sr}$ could lead to studies of low-entropy states in the Fermi-Hubbard model with SU(N) symmetry~\cite{cazalilla2014ultracold}, which could also be extended to alkalis with tunable Feshbach resonance using Raman sideband cooling~\cite{kaufman2014two,parsons2015site}.
Additionally, the tunable density of this approach combined with long-range-interacting particles could be used to prepare lower-density supersolid ground states of extended Hubbard models, while circumventing inelastic losses that can arise during evaporative cooling of such species, like molecules~\cite{baier2016extended, guardado2021quench, chomaz2022dipolar, su2023dipolar, christakis2023probing, weckesser2024realization, carroll2025observation}. And, in all cases --- bosons, fermions, and long-range interacting systems --- one might combine atomic rearrangement with entropy redistribution for fast, ultralow-entropy preparation, since the starting point of entropy redistribution is typically an incompressible, gapped many-body state, e.g. a band- or Mott-insulator, which is accessible via atomic rearrangement~\cite{mazurenko2017cold,xu2025neutral}. 
Finally, the combination of programmability, ultracold single-particle preparation, and optical lattices demonstrated here may be a powerful paradigm for fermionic quantum processors~\cite{gonzalez2023fermionic,ott2025error}.

\begin{acknowledgments}

The authors wish to thank A.~Carroll, R.~Kaubruegger, and P.~Preiss for careful readings of the manuscript and helpful comments. 
We acknowledge helpful discussions with  M.~Greiner,  C.~Regal, J.~Simon, D.~S.~Weiss, R.~Wang and J.~Zeiher. We are also grateful to N.~Schine and M.~A.~Norcia for early contributions to the experiment.
This material is based upon work supported by the Army Research Office (W911NF-22-1-0104), the U.S. Department of Energy, Office of Science, National Quantum Information Science Research Centers, Quantum Systems Accelerator, Physics Frontier Center PHY-2317149, and the National Institute of Standards and Technology. 
We also acknowledge funding from Lockheed Martin.
A.C. acknowledges support from the NSF Graduate Research Fellowship Program (Grant No. DGE2040434); W.J.E. acknowledges support from the NDSEG Fellowship; N.D.O. acknowledges support from the Alexander von Humboldt Foundation.
\end{acknowledgments}

\bibliography{references.bib}

\clearpage

\setcounter{figure}{0}
\renewcommand{\figurename}{Extended Data Fig.}

\section*{Methods}

\subsection*{Image analysis}\label{sec:image_analysis}

We infer the locations of atoms in the optical lattice by analyzing the second and third images (see Fig.~\ref{fig:1}) from each run of the experiment. The second image is taken after running a rearrangement protocol, as described in~\cite{young2024atomic}. To track the spatial phase of the lattice, we compare the first or second image to a simulated image with Gaussian masks placed at each location in the target pattern. By scanning the $x$ and $y$ offsets of the Gaussian masks to maximize the pixel-wise product of the simulated image with the experimentally observed image, we can infer the phase of the optical lattice relative to the camera sensor.

With knowledge of the spatial phase of the lattice on the camera sensor, it is possible to identify the discrete set of locations where atoms might be located. Given the size of our field of view, we construct a set of $85\times85 = 7225$ lattice site locations. The distribution of photons from a given lattice site on the camera sensor is given by the point spread function (PSF) of the imaging system. By assuming a Gaussian PSF whose width is constant across the system, we can construct a simple model for how an underlying distribution of photons emitted from each lattice site will be distributed on the camera sensor. Suppose $u_i$ describes the true number of photons emitted into the imaging system from lattice site $i$. The expected number of photon detections on pixel $j$ -- denoted $d_j$ -- will be given by $d_j = \sum_i P(|\vec{r}_i - \vec{r}_j|) \, u_i$. Here $P$ is the PSF, and $\vec{r}_{i}$ is the location of lattice site $i$.

In practice, fluorescence images give a measurement of the quantity $d_j$. In order to invert this model and develop an estimate of $u_i$, we run the function `scipy.sparse.linalg.lsqr'~\cite{2020SciPy-NMeth} to solve for a least-squares solution to the linear equations 
\begin{align}
    \sum_{i, j} \left[ P\left(|\vec{r}_i - \vec{r}_j|\right) \, u_i - d_j \right] = 0.
\end{align} 
By thresholding on the inferred value of $u_i$, we determine whether we believe there to be an atom at the lattice site $i$.

\subsection*{Postselection}\label{sec:post_sel}

Data are post-selected based on the rearranged images.
For all data, we reject experimental trials in which an atom is detected outside of the target rearrangement pattern.
For all data with $N=4, 18$, we reject trials for which fewer than $N$ atoms are prepared in the $N$-site target pattern.
For data with $N=50$ we reject trials in which fewer than 45 atoms are prepared in the target pattern.

The images in the top row of Fig.~\ref{fig:2} (read from left to right) are averaged over $935, 958, 941, 920, 1000$ experimental trials.
The images in the bottom row of Fig.~\ref{fig:2} (read from left to right) are averaged over $3401, 3448, 3514, 3422, 3481$ experimental trials.
The data used to infer temperatures of data with $(N, V_L / E_R) = (4, 5.1), (18, 5.1), (50, 5.1)$ [shown in Fig.~\ref{fig:3}(a)] have $3258, 935$ and $241$ trials, respectively.
Finally, the data in Fig.~\ref{fig:3}(d) with $N=4, 18, 50$ are averaged over $5853, 3401, 773$ trials, respectively.
We emphasize that all trials in these averages satisfy the corresponding postselection criteria described above.

\subsection*{Sideband cooling}
\begin{figure}
    \includegraphics[width=\columnwidth]{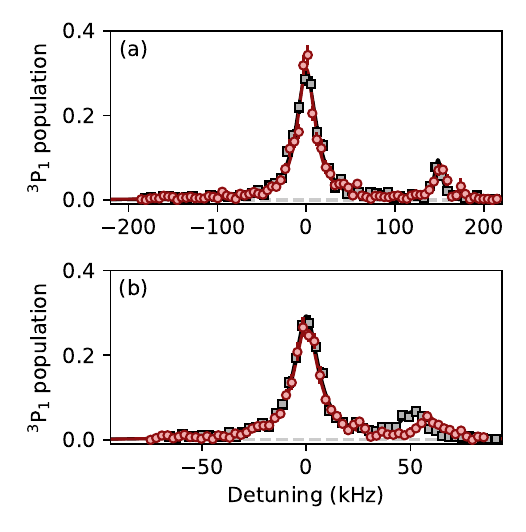}
    \caption{\label{fig:1ED}%
    \textbf{Sideband thermometry before and after Hubbard state preparation.}
    Measured excitation probability of ${}^3{\rm{P}}_1$ atoms for spectroscopy performed before (black squares) and after (red circles) the Hubbard state preparation protocol.
    Panel (a) [(b)] shows the spectrum for one of the in-plane directions [the out-of-plane direction].
    The ``before" thermometry probes the initially stochastically filled ensemble shown in the first panel of Fig.~\ref{fig:1}(c).
    The ``after" thermometry is taken by preparing an $N=50$ ensemble at $20.8(3)\,E_R$, and then ramping both lattices up quickly to pin the atoms while remaining adiabatic with respect to the band gap.
    Solid lines are fits to a sum of three Lorentzians, with the amplitudes $A_{\rm rsb}$ ($A_{\rm bsb}$) of the red (blue) sidebands related to the inferred mean motional occupation number $\bar{n}$ by the relation $\bar{n} = 1/(A_{\rm bsb}/A_{\rm rsb}-1)$. 
    }
\end{figure}

After the second image, we perform resolved sideband cooling with atoms in the optical lattice, the tweezers turned off, and on the 7.5-kHz-wide ${}^1{\rm{S}}_0\leftrightarrow{}^3{\rm{P}}_1$ transition.
The trap frequencies during sideband cooling are $162.4(7)\,\rm{kHz}$, and $148.0(6)\,\rm{kHz}$ in the $x$-$y$ plane, and $55.8(4)\,\rm{kHz}$ in the $z$-direction.
Sideband thermometry in the lattice~\cite{young2022tweezer} suggests mean single-atom motional occupation numbers of $\overline{n}_x = 0.045(21)$, $\overline{n}_y = 0.045(25)$, $\overline{n}_z = 0.13(3)$ along the two in-plane directions and the $z$ direction, respectively.
Along with the equation for the thermal entropy in a 3D quantum harmonic oscillator,
\begin{align}
    S_{\rm QHO}^{\rm (3D)}(\overline{n}_x, \overline{n}_y, \overline{n}_z) = k_B \sum_{\alpha \in \{x,y,z \}} &(1 + \overline{n}_\alpha) \ln\left(1 +  \overline{n}_\alpha \right) \nonumber \\
    &- \ \overline{n}_\alpha \ln\left(\overline{n}_\alpha\right),
\end{align}
these values are used to calculate the expected contribution to $S/N$ due to imperfect laser cooling, as plotted in Fig.~\ref{fig:4}(a).

To check for potential heating to higher bands during our state preparation protocol which could break the tight-binding assumption of our theoretical modeling, we additionally perform a comparison of the sideband thermometry before and after the Hubbard state preparation in Extended Data Fig.~\ref{fig:1ED}.
Within the precision of this thermometry technique, there is no observable heating of the sample either in-plane or out-of-plane.
There is an observable increase in the sideband frequency in the spectra after the state preparation due to the density distribution of the atoms dominantly sampling the central, deepest part of the lattices; this is in contrast to the thermometry before state preparation which effectively averages over a large $48\times48$ site region.

\subsection*{Characterization of lattice shapes}

The procedures for characterizing the shapes of the pancake trap and 2D lattice both employ spectroscopy on the ${}^1\rm{S}_0 \leftrightarrow {}^3\rm{P}_1$ electronic transition, and begin by stochastically loading ${}^{88}\rm{Sr}$ atoms into a $16 \times 24$, $515\,\rm{nm}$ tweezer array, with inter-tweezer spacings of $3\,a_L$ and $2\,a_L$ along the $x$ and $y$ axes, respectively. 
The magnetic fields are tuned to the magic angle for the ${}^1\rm{S}_0 \leftrightarrow {}^3\rm{P}_1$ transition with atoms in $515\,\rm{nm}$ optical tweezers~\cite{norcia2018microscopic}.
These magnetic fields do not satisfy a magic condition for the $813.427\,\rm{nm}$ lattice light, which thereby induces a light shift on the ${}^1\rm{S}_0 \leftrightarrow {}^3\rm{P}_1$ transition. 
For the tweezer and lattice depths used in these measurements, the magnitude of this shift is linear in the intensity of the lattice light. 
Therefore, by measuring the shifts of the ${}^1\rm{S}_0 \leftrightarrow {}^3\rm{P}_1$ transition induced by the axial lattice at each tweezer location, we can infer the relative intensity across the array.
To characterize the shape of the 2D lattice, we perform a similar procedure. 
However, after ramping up the 2D lattice, we additionally ramp the axial lattice up to a depth of $V_0^{({\rm{Ax}})} / \hbar = 2\pi \times 12.0(2) \, \rm{kHz}$, and then ramp the tweezers off. 
This ensures that atoms sit in the potential wells of the 2D lattice during spectroscopy, even if the tweezers are not well referenced to the lattice phase. 
The axial lattice is necessary for confinement along the $z$-axis, and is shallow enough that it induces a negligible level of inhomogeneous broadening. 

After collecting spectroscopy data, we fit the ${}^1\rm{S}_0 \leftrightarrow {}^3\rm{P}_1$ resonances at each tweezer position to a model for a rotated 2D Gaussian,
\begin{align}
    f(x,y) = f_0 \, e^{-\left[a(x - x_0)^2 + 2b(x - x_0)(y - y_0) + c(y - y_0)^2\right]},
\end{align}
where
\begin{align}
    a &= 2\frac{\cos^2\left(\theta\right)}{w_u^2} + 2\frac{\sin^2\left(\theta\right)}{w_v^2}, \\
    b &= -2\frac{\sin\left(\theta\right)\cos\left(\theta\right)}{w_u^2} + 2\frac{\sin\left(\theta\right)\cos\left(\theta\right)}{w_v^2}, \\
    c &= 2\frac{\sin^2\left(\theta\right)}{w_u^2} + 2\frac{\cos^2\left(\theta\right)}{w_v^2}, 
\end{align}
$\theta$ is the angle of the Gaussian relative to the $x$-$y$ coordinate axes of the lattice, $w_u, w_v$ are the Gaussian waists along the principal axes of the elliptical 2D Gaussian, and $f_0$ is a proportionality constant. 
For the axial lattice we find $[w_u, w_v, \theta] = [24.7(1){\,\rm{\upmu m}}, 36.4(1){\,\rm{\upmu m}}, -14.9^{\circ}(1^{{\circ}})]$.
For the 2D lattice we find $[w_u, w_v, \theta] = [90(6){\,\rm{\upmu m}}, 80(7){\,\rm{\upmu m}}, -41^{\circ}(1^{{\circ}})]$.

\subsection*{Calibration of lattice depths}

To calibrate the depth of the axial lattice, we ramp ${}^{88}\rm{Sr}$ atoms from the tweezers into the pancake trap as described in the main text, and shown in Fig.~\ref{fig:1}(a). 
We then modulate the intensity of the axial lattice, which forms the pancake trap. 
When the frequency of this modulation is twice the harmonic trapping frequency, this modulation causes parametric excitation and eventually loss of atoms from the trap~\cite{blatt2015low}. 
To determine the trap frequencies of the axial lattice potential, we scan the frequency of the applied intensity modulation, and measure the fraction of atoms that are lost. 
Knowledge of the trap frequency allows one to calibrate the depth of a potential $V(x,y,z)$ with the form
\begin{align}
    V(x) = - V_0 \, e^{-2x^2 / w^2}
\end{align}
by the relation
\begin{align}
    V_0 =  \frac{1}{4} m \omega^2 w^2,
\end{align}
where $V_0$ is the depth, $m$ is the atomic mass, $\omega$ is the angular trap frequency, and $w$ is the Gaussian waist of the potential, as calibrated by the procedure described in \textit{Characterization of lattice shapes}.
For the depth used during adiabatic state preparation, we find that the two in-plane trap frequencies (vertical trap frequency) of the pancake trap are $94.8(9)\,\rm{Hz}$ and $63.2(9)\,\rm{Hz}$ [$3950(10)\,\rm{Hz}$].

To calibrate the depth of the 2D lattice, we perform single-particle quantum-walk experiments at three different depths and multiple evolution times~\cite{young2022tweezer}. By repeating these experiments many times and comparing the observed density distribution to a tight-binding model for single-particle walks in a flat bow-tie lattice, we can estimate the depth of the lattice~\cite{young2023programmable}. 

\subsection*{Calibration of Hubbard parameters}

Values of the hopping parameter $J$ quoted in the main text are calculated at each lattice depth with a single-particle band-structure calculation, which yields the ground-band Wannier functions $w_i(\vec{r})$ for lattice site $i$.
The equation for $J$ is
\begin{equation}
    \hbar J = - \int \,\text{d}\vec{r} \, w^*_0(\vec{r}) \left[ -\frac{\hbar^2}{2m} \nabla^2 + V(\vec{r}) \right]\, w_j(\vec{r}) 
\end{equation}
where $i=0$ indexes the central lattice site, $i=j$ an adjacent site; $V(\vec{r})$ is the lattice potential.
Next, for the interaction parameter $U$, we use values calculated with the equation
\begin{equation}
    \hbar U = \frac{4 \pi \hbar^2 a_{sc}}{m} \int \text{d}\vec{r} \, |w_0(\vec{r})|^4 \label{eq:U_pert}
\end{equation}
where $a_{sc}$ is the s-wave scattering length~\cite{jaksch1998cold}. 
The finite extent of the lattices means that the axial confinement and local lattice depth vary spatially.
This changes the shape of $w_i(\vec{r})$ for different $i$ and in principle $J$ and $U$ also vary with $i$.
However, states are not large compared to the lattice sizes discussed in \textit{Characterization of lattice shapes}, and we take $J$ and $U$ to be homogeneous.

At the five 2D lattice depths $V_L \in \{ 5.1(1), 9.0(1), 12.9(1), 16.9(2), 20.8(3) \} E_R$, Eq.~\ref{eq:U_pert} predicts values of approximately $U \in 2\pi \times \{ 990, 1500, 2000, 2400, 2700 \} \, \rm{Hz}$, respectively.
Eq.~\ref{eq:U_pert} is valid when $\omega_z \gg U$, where $\omega_z = 2\pi \times 3950\,\rm{Hz}$ is the vertical trapping frequency of the pancake trap.
As this condition breaks down, higher bands in the lattice need to be considered.
To estimate the error in the values of $U$, we perform an exact calculation for the two-particle interaction strength $U_{\rm QHO}$ in a 3D harmonic trap~\cite{Idziaszek2006analytical}, with trap frequencies calculated from the harmonic approximation of a lattice potential well.
We view $U_{\rm QHO}$ as a lower bound on the true interaction strength, and find that $0.75 \, U < U_{\rm QHO} < 0.85 \, U$ for all depths quoted in the main text.
For $V_L \in \{9.0(1), 12.9(1), 16.9(2), 20.8(3) \} \, E_R$, this level of error has little impact on the physics, as the interaction strength is so much larger than the hopping energy $J$ that double occupancies are almost completely suppressed---even with a $25\%$ reduction in $U$.
For $V_L = 5.1(1) \, E_R$, we repeat the thermometry procedure with an interaction strength of $0.8 \, U$ and find that the inferred inverse temperature $\beta$ does not change by more than the $\pm 10 \%$ error bar.

The chemical potential $\mu_i$ at a given lattice site $i$ can be written $\mu_i = \mu_0 + \Delta\mu_i$.
Here, $\mu_0$ is a global offset which -- in QMC calculations -- is chosen to match the simulated atom number to the experimental value.
Due to the harmonic confinement provided by the optical lattices, $\mu_i$ varies from site to site, and we model this variation with $\Delta\mu_i$.
The terms $\Delta\mu_i$ are inputs to mean-field and QMC simulations, and must be characterized independently. 

Physically, the parameter $\Delta\mu_i$ represents the energy difference between having an atom at lattice site zero (where $\mu_{i = 0} = \mu_0$) and lattice site $i$.
This energy difference arises from two effects. 
First, one must consider the difference between the potential-energy offset between $i$ and site $0$.
This can be determined from the results of procedures described in \textit{Characterization of lattice shapes} and \textit{Calibration of lattice depths}.
Second, we take into account the variation in zero-point energy for atoms localized at lattice site $i$ versus site $0$.
For example, in the deep-lattice limit, one can approximate these zero-point energies by the ground-state energies for an atom in a 3D harmonic oscillator $\hbar \left( 2 \omega_{\rm rad} + \omega_{\rm Ax}\right) / 2$, where $\omega_{\rm rad}$ is the angular trap frequency for lattice site $i$ in the $x-y$ plane, and $\omega_{\rm Ax}$ is the angular trap frequency along the $z$ direction, which can vary with lattice site.
In deep lattices, the ground band is approximately flat, and there is almost no ambiguity about what the trap frequency is.
In shallow lattices, the ground-band energy varies significantly across different quasimomenta.
In order to estimate the effective ground-state energy of a localized atom in the Wannier function for site $i$, we use the mean energy of the ground band across all quasimomenta.

\subsection*{Mean-field model}

While quantitatively inexact, mean-field (MF) models have provided insight into the qualitative features of the Bose-Hubbard model~\cite{sheshadri1993superfluid, sheshadri1995percolation, jaksch1998cold, van2001quantum, pai2012bose}. To calculate the $T=0$ superfluid density shown in Fig.~\ref{fig:1}(d), we construct an inhomogeneous MF model, following the procedure presented in~\cite{sheshadri1995percolation}. 
The central approximation is to take
\begin{align}
    \hat{b}_i^\dagger \hat{b}_j \approx \langle \hat{b}_i^\dagger \rangle \hat{b}_j + \hat{b}_i^\dagger \langle \hat{b}_j \rangle - \langle \hat{b}_i^\dagger \rangle \langle \hat{b}_j \rangle .
\end{align}
Following~\cite{sheshadri1995percolation, pai2012bose}, we define the parameters
\begin{align}
    \psi_i &= \langle \hat{b}_i \rangle \\
    \phi_i &= \frac{1}{z} \sum_\delta \psi_{i+\delta},
\end{align}
where $\delta$ labels the $z$ nearest-neighbor lattice sites around site $i$. The MF Bose-Hubbard Hamiltonian then decouples into a sum over local terms, and the ground state can be be found with an iterative, self-consistent procedure~\cite{pai2012bose}.

\subsection*{Quantum Monte Carlo simulations}

We performed path integral quantum Monte Carlo simulations using the worm algorithm~\cite{Prokofev1998} in the formulation of Ref.~\cite{Sadoune2022} where use was made of the ALPSCore libraries~\cite{ALPSCore}. 
Simulations were performed in the grand-canonical ensemble. The chemical potential was fine-tuned in order to reproduce the target particle number (4, 18, and 50) on average. 
The simulations took into account nearest-neighbor hopping with amplitude $J$ and on-site interactions with amplitude $U$, as described in \textit{Calibration of Hubbard parameters}. 
Various observables were computed such as the in-situ density distribution, the density fluctuations, the energy, and the equal-time density matrix, among others. Scans over different temperatures were performed in order to perform thermometry on the density distribution. Entropies were computed through thermodynamic integration of the energy $\tilde{E} = \left< E \right> - \mu \left< N \right>$ using the formula
\begin{equation}
S(\beta_1) - S(\beta_2) = \beta_1 \tilde{E}(\beta_1) - \beta_2 \tilde{E}(\beta_2) + \int_{\beta_1}^{\beta_2} \tilde{E}(\beta') d\beta'.
    \label{eq:entropy_integration}
\end{equation}
A fine grid of inverse temperatures between a high-temperature cutoff (see below) and the target temperature with spacing $\Delta \beta = 0.1/J$ was chosen, and we fitted a cubic spline through the corresponding energies before performing the integration.
At high temperature, in general taken to be $\beta J = 0.1$, the thermodynamic properties were obtained from the high-temperature series expansion to second order, where the free energy density is given by
\begin{equation}
    \Omega/N = -T \ln f_0 - T d f_2,
\end{equation}
where the dimension $d=2$, the zeroth order term, $f_0$, is
\begin{equation}
    f_0 = \sum_{n=0}^{\infty} e^{\mu \beta n - \frac{U}{2} \beta n(n-1)},
\end{equation}
and the second order correction reads
\begin{eqnarray}
    f_2 & = & \left( \frac{J}{f_0 U} \right)^2 \sum_{n_1, n_2=0}^{\infty} e^{\mu \beta (n_1 + n_2) - \frac{U}{2} \beta [n_1(n_1-1) + n_2(n_2-1)] } \times \nonumber \\
 {} & {} & \left[ \frac{ n_1(n_2+1)  g[U \beta (n_1 - n_2 -  1)]}  {(n_1 - n_2 - 1)^2} + (n_1 \leftrightarrow n_2) \right].
\end{eqnarray}
The function $g(x)$ is given by $g(x) = e^{x} - x -1 $.
The entropy thus obtained at $\beta J = 0.1$, in combination with the local density approximation, served as the high-temperature cutoff of Eq.~\ref{eq:entropy_integration}.

\subsection{Ideal time-of-flight expansions}

After a non-interacting, 2D, free-space expansion in the far-field limit and with time $t_{\rm TOF}$, the atomic density $\tilde{n}(k_x, k_y)$ can be expressed
\begin{align}
    \tilde{n}(k_x, k_y) &= \left(\frac{m}{\hbar t_{\rm TOF}}\right)^2 \left|\tilde{w}\left(\vec{k}  \right)\right|^2 \mathcal{S}\left( \vec{k} \right), \\
    \mathcal{S}\left( \vec{k} \right) &= \sum_{i,j} e^{i \vec{k} \cdot \left( \vec{r}_i - \vec{r_j} \right)} G^{(1)}_{ij}\label{eq:struct}
\end{align}
where $m$ is the mass of a single atom, $G^{(1)}_{ij} = \langle \hat{b}_i^\dagger \hat{b}_j \rangle$ and  $\vec{k} = m \vec{r} / \hbar t_{\rm TOF}$.
The coherence function $G^{(1)}_{ij}$ is sensitive to the phase.
In particular, the observation of diffraction patterns with resolvable peaks demonstrates the presence of a coherence length that is comparable to the size of the system.
More specifically, large variations in the phase across the system would lead to an incoherent summation in Eq.~\ref{eq:struct}, and the interference pattern could not exhibit structured diffraction peaks~\cite{gerbier2008expansion}.
Phase coherence is central to the phenomenon of superfluidity~\cite{fisher1973helicity}.
However, it is important to distinguish the existence of phase coherence from the presence of superfluidity.
For example, even in a MI phase, TOF patterns can exhibit low-contrast diffraction peaks (indicative of short-ranged phase coherence) due to particle-hole fluctuations~\cite{gerbier2005phase}.
Therefore, we take the QMC-inferred `ideal' TOF expansions in Fig.~\ref{fig:3}~(c) as indications of phase coherence, but not, on their own, evidence for superfluidity.

\subsection{Semi-classical time-of-flight simulation}

In the main text, we note that TOF expansions are performed in the presence of the pancake trap, leading to harmonic confinement that modifies the expansion.
In order to simulate the effect on the dynamics, we employ a semi-classical approach, simultaneously sampling the QMC-calculated position and momentum distributions that best represent experimental conditions, inferred based on the thermometry procedure described in the main text.
After sampling the initial positions and momenta for a given number of atoms, the state of each atom is evolved independently (i.e. ignoring interactions) for the experimental expansion time of $1.5\,\rm{ms}$. 
The simulation employs classical equations of motion, and takes into account the harmonic confinement of the pancake trap.
To model our detection protocol, the final position of each atom is taken to be the location of the nearest lattice site, which allows us to account for the parity projection inherent to our imaging.

\subsection{Error bars}

The dominant source of uncertainty in our work is thermometry. We conservatively estimated that the fitting discrepancy between QMC data and experimental data results in a 10\% uncertainty on the temperature: Deviations between the QMC data over such a temperature range correspond to the scale of error bars and fluctuations of the experimental data. Therefore, all error bars in this work derive from $\pm 10\,\%$ error bars for $\beta$, and other sources such as {\it e.g.}, the error bars and the choice of temperature grid in the entropy calculation, are negligible.

\subsection{Characterizing atom loss}

Here we investigate the magnitude of atom loss present in the data.
The final atom number observed is impacted by various effects including actual atom loss during the adiabatic state preparation, loss during imaging, parity projection during imaging, and infidelity in the determination of occupied and empty lattice sites.
Though it is not feasible to robustly extract all of these quantities separately from the presented data, we can isolate some of these effects using the following procedure.
In particular, we fit the final observed atom number distribution assuming the initial atom number distribution undergoes three stochastic processes in sequence: loss per atom with probability $L$, pairwise loss per atom pair with probability $P$ and additional atom detection per atom with probability $F$.
While ideally we would like to determine the loss occurring solely due to the many-body state preparation, in practice the fitted loss per atom also includes effects such as imaging loss and false negative detections which are not easily distinguished by this method.
The pairwise loss stage is intended to capture the effect of parity projection from doubly-occupied sites in the many-body state; unlike the other two stages, the fitted pairwise loss should be interpreted as the chance that any possible pair of atoms undergoes this pairwise loss as opposed to any individual atom.
Finally, the additional atom detection stage is intended to capture the rate of false positive detections induced by the leakage of fluorescence of an actual atom on one site to its neighboring sites.

To perform the fit, we use maximum likelihood estimation.
From the measured initial atom number distribution and given the parameters $\theta = \lbrace L,P,F \rbrace$, we calculate the resulting multinomial distribution characterized by $\lbrace p_n(\theta) \rbrace$ for observing $n$ atoms in the final image on any given trial.
Our measurements are described by the set $M = \lbrace m_n \rbrace$, where $m_n$ is the number of trials with $n$ atoms observed in the final image.
The log-likelihood function, ignoring constant terms that do not depend on the model parameters $\theta$, is then given by $\mathrm{log} \, \mathcal{L}\left(\theta | M \right) \sim \sum_n m_n \mathrm{log} \, p_n(\theta)$.
Maximizing this quantity yields the maximum likelihood estimate of the parameters $\theta$, and we characterize the error in these estimates by bootstrap resampling.
For $N=4$ [$N=18$], the fitted loss per atom $L$ varied between $0.02-0.035$ [$0.03-0.05$] across the full range of lattice depths $V_L$ shown in Fig.~\ref{fig:2}.
For the $N=50$ data, $L$ appears to be $\lesssim 0.1$; however, we note that for these data the fits are highly uncertain with the error in $L$ being comparable to its magnitude, likely due to a combination of the lower number of trials, the larger state space of outcomes and the spread in initial atom number.

\end{document}